\documentclass[10pt,aps,prl,twocolumn]{revtex4-1}
\usepackage{graphicx}
\usepackage{amssymb}
\usepackage{amsmath}
\usepackage{appendix}
\usepackage{comment}

\newcommand{\eq}[1]{\begin{equation}#1\end{equation}}

\usepackage{mdframed}

\begin{document}

\title{Generalized Boundary Conditions for Spin Transfer}

\author{Yaroslav Tserkovnyak and Hector Ochoa}
\affiliation{Department of Physics and Astronomy, University of California, Los Angeles, California 90095, USA}

\begin{abstract}
We develop a comprehensive description of static and dynamic spin-transfer torque at interfaces between a normal metal and a magnetic material. Specific examples of the latter include ferromagnets,  collinear and noncollinear antiferromagnets, general ferrimagnets, and spin glasses. We study the limit of the exchange-dominated interactions, so that the full system is isotropic in spin space, apart from a possible symmetry-breaking order. A general such interface yields three coefficients (corresponding to three independent generators of rotations) generalizing the well-established notion of the spin-mixing conductance, which pertains to the collinear case. We develop a nonequilibrium thermodynamic description of the emerging interfacial spin transfer and its effect on the collective spin dynamics, while circumventing the usual discussion of spin currents and net spin dynamics. Instead, our focus is on the  dissipation and work effectuated by the interface. Microscopic scattering-matrix based expressions are derived for the generalized spin-transfer coefficients.
\end{abstract}

\maketitle

\textit{Introduction.}|The problem of interfacial spin transfer, along with the associated spin torque \cite{slonczewskiPRB89,*slonczewskiJMMM96,*bergerPRB96} and spin pumping \cite{tserkovPRL02sp,mizukamiJJAP01,*urbanPRL01,*heinrichPRL03}, has been central to the field of metal-based spintronics for over twenty years \cite{tserkovRMP05,ralphJMMM08}. For much of its history, the focus has  been on the dynamics of collinear ferromagnets. In this case, the spin-mixing conductance has become the key quantity for describing both the spin torque \cite{brataasPRL00} and the spin pumping \cite{tserkovPRL02sp}, which have subsequently being recognized as Onsager-reciprocal processes \cite{tserkovPRB08mt,brataasCHA12}. Recently, a straightforward generalization to the dynamics of collinear antiferromagnets has been put forward \cite{takeiPRB14,chengPRL14}. In particular, it has been argued \cite{baltzCM16} that at frequencies much smaller than the exchange energy, the interfacial spin transfer is dominated by the rigid N{\'e}el-order dynamics. As such, it can be parametrized by an antiferromagnetic spin-mixing conductance \cite{takeiPRB14}, in close analogy to the ferromagnetic case, yielding only small corrections due to the internal canting dynamics.

\begin{figure}[t]
\includegraphics[width=0.9\linewidth]{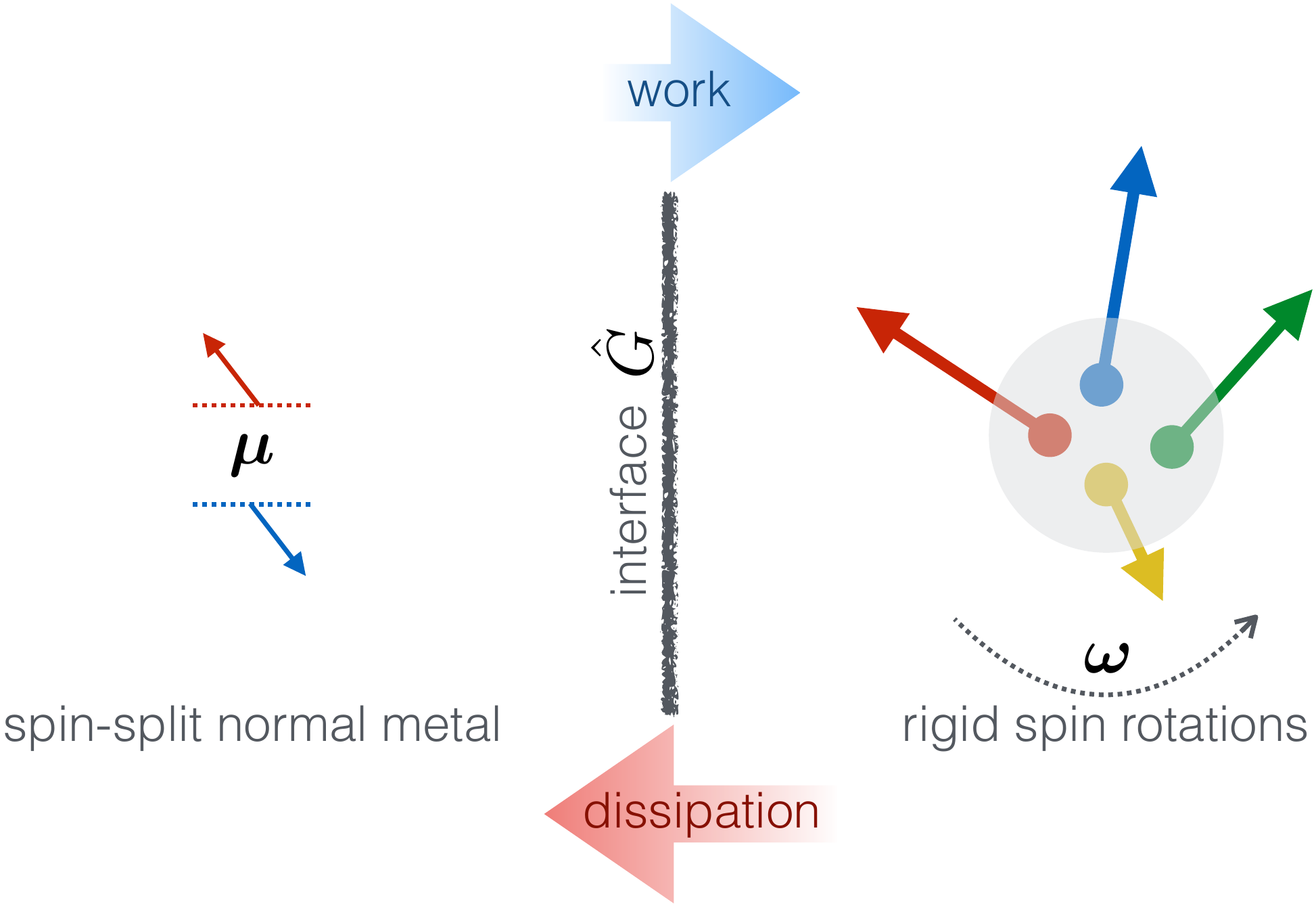}
\caption{A schematic of the magnetic system (right) in contact with a normal metal (left). The nonequilibrium spin state of the metal is parametrized by the (vectorial) spin accumulation $\boldsymbol{\mu}$. The magnet, whose spin arrangement is determined by some isotropic exchange Hamiltonian, is described, near the interface ($x=0$), by uniform (rigid) rotations of all spins. Its instantaneous nonequilibrium state is thus characterized by the (vectorial) frequency of SO(3) rotation $\boldsymbol{\omega}$. The $3\times3$ matrix $\hat{G}$, which is governed by the electron reflection amplitudes at the interface, generalizes the concept of the spin-mixing conductance pertinent to the collinear case. The central object of the theory is the modified Rayleigh dissipation function \eqref{R}, expressed in terms of $\hat{G}$, $\boldsymbol{\omega}$, and $\boldsymbol{\mu}$.}
\label{fig1}
\end{figure}

In this Letter, we generalize the description of the low-frequency torque and pumping to noncollinear magnetic configurations. The main underlying assumption is that the interactions near the interface are dominated by the spin-isotropic exchange coupling (of arbitrary form, allowing, in particular, for frustration). At low frequencies, the associated spin dynamics near the interface can be captured in terms of rigid SU(2) rotations, with the spin-mixing conductance generalized to a $3\times3$ positive-definite matrix. (See Fig.~\ref{fig1} for a schematic.) The latter, when diagonalized along certain principal axes locked to the magnet's spin rotations, can be parametrized by three independent coefficients. The theory naturally lends itself to noncollinear antiferro- and ferrimagnets as well as spin glasses \cite{halperinPRB77,andreevSPU80,gomonayPRB12}.

We argue that the most streamlined description of spin transfer in this generalized setting is accomplished by departing from the usual analysis of the interfacial spin currents and instead focusing on energy. Namely, the central object of the theory is the Rayleigh dissipation function for the magnetic heat pumping into the normal metal, offset by the appropriate work on the collective magnetic dynamics (either ordered or disordered) by the spin-transfer torque. Our perspective is thus based on energetics rather than spin conservation (albeit the latter is recovered in the appropriate cases). Following a general construction, we will check the new methodology against the known spin-torque/pumping results for the collinear (anti)ferromagnets, and then apply it to the case of spin glasses.

\textit{Phenomenology.}|The collective magnetic dynamics near the interface are parametrized as a rigid rotation of spins. This corresponds to the low-frequency limit, when all the relevant energy scales in the magnet (associated with anisotropies, Dzyaloshinsky-Moriya interactions, magnetic field, as well as the driving frequency) are much lower than the microscopic exchange interaction. In this limit, the largest-amplitude dynamics correspond to the spin rotations as a whole, along with smooth spatial textures thereof \cite{baltzCM16}. The latter are inconsequential to our interfacial analysis. For simplicity, we start by assuming the magnet is insulating.

At low frequencies, the instantaneous dissipation rate associated with the magnetic dynamics depicted in Fig.~\ref{fig1} can generally be written as \cite{landauBOOKv5,Note0}
\eq{
P=\boldsymbol{\omega}^T\hat{G}\boldsymbol{\omega}=\omega_iG_{ij}\omega_j\,,
\label{P}}
summing over the repeated indices. $\hat{G}$ is a positive-definite (symmetric) $3\times3$ matrix parametrizing heat flow into the normal metal, which (microscopically) depends on the strength of the exchange coupling across the interface. The (spin) frame of reference can be rotated to diagonalize $\hat{G}\to\{G_1,G_2,G_3\}$, where $G_i\geq0$ are the (generally) anisotropic damping parameters for rotations about the corresponding (principal) axes.

The usual Rayleigh dissipation function would be given by half of the dissipation power \eqref{P} \cite{landauBOOKv5}. In the presence of a spin accumulation $\boldsymbol{\mu}$ in the metal, however, the interfacial energy flow gets modified, due to the work done by $\boldsymbol{\mu}$ on the magnetic system \cite{kimPRB15br}. In the special case of $\boldsymbol{\mu}=\hbar\boldsymbol{\omega}$, in particular, we see, from the corotating frame of reference, that the combined system is in the state of a mutual equilibrium \cite{tserkovPRB05}. Indeed, the spin accumulation is cancelled by the fictitious field $\hbar\boldsymbol{\omega}$ due to the rotation, while the spins in the magnet are static. In this case, the electrons of the metal should not exert any torque on the magnetic dynamics. The correspondingly modified Rayleigh dissipation function, which accounts for the work done by the spin-accumulation induced torque, is thus deduced to be
\eq{
R=\frac{1}{2}\tilde{\boldsymbol{\omega}}^T\hat{G}\tilde{\boldsymbol{\omega}}\,,
\label{R}}
where $\tilde{\boldsymbol{\omega}}\equiv\boldsymbol{\omega}-\boldsymbol{\mu}/\hbar$ vanishes in the aforementioned state of the mutual (dynamic) equilibrium \cite{Note1}. We will now develop a microscopic, scattering-matrix based theory for calculating $\hat{G}$, before applying Eq.~\eqref{R} to some concrete examples of the (Lagrangian) magnetic dynamics.

\textit{Scattering-matrix formalism.}|To introduce the relevant microscopic concepts in the simplest setting, we start with the case of a single quantum transport channel in the normal metal. The reflection matrix thus has dimensions $2\times2$: 
\eq{
\hat{r}\equiv\{r^{\sigma\sigma'}\}\,,
}
with $r^{\sigma\sigma'}$ standing for the interfacial electron scattering coefficients for spin $\sigma'$ into $\sigma$. Having obtained the reflection matrix in a certain (spin) frame of reference at time $t=0$, it would become
\eq{
\hat{r}(t)=\hat{U}(t)\hat{\mathfrak{r}}\hat{U}^\dagger(t)\,,
\label{r}}
at a later time $t$ [denoting $\hat{r}(0)$ by $\hat{\mathfrak{r}}$]. The time-dependent SU(2) transformation $\hat{U}(t)$ describes the instantaneous state of the magnet, corresponding to a three-dimensional rotation of the ($t=0$) reference state. The equation of motion for the rotation matrix is
\eq{
i\hbar\frac{d}{dt}\hat{U}(t)=\boldsymbol{\omega}(t)\cdot\mathbf{\hat{s}}\,\hat{U}(t)\,,
}
with the initial condition of $\hat{U}(0)=1$. $\mathbf{\hat{s}}$ is the electron spin operator (i.e., $\hbar/2$ times a vector of the Pauli matrices $\boldsymbol{\hat{\sigma}}$) and $\boldsymbol{\omega}$ is the (vectorial) angular velocity.

The energy dissipation rate, for a given instantaneous frequency of rotation $\boldsymbol{\omega}$, is given by \cite{moskaletsPRB02dn,brataasPRL08}
\eq{
P=\frac{\hbar}{4\pi}{\rm Tr}\left[\dot{\hat{r}}\dot{\hat{r}}^\dagger\right]\,,
\label{Prr}}
where $\dot{\hat{r}}\equiv d\hat{r}/dt$ is the rate of change of the reflection matrix. Substituting Eq.~\eqref{r} into \eqref{Prr}, we find
\eq{
P=\frac{\hbar}{8\pi}{\rm Tr}\left[\boldsymbol{\omega}^2-\hat{\mathfrak{r}}(\boldsymbol{\underline{\omega}}\cdot\boldsymbol{\hat{\sigma}})\hat{\mathfrak{r}}^\dagger(\boldsymbol{\underline{\omega}}\cdot\boldsymbol{\hat{\sigma}})\right]\,,
}
where $\boldsymbol{\underline{\omega}}\equiv \hat{R}^{-1}\boldsymbol{\omega}$, $\hat{R}$ being the SO(3) rotation matrix corresponding to the SU(2) spin rotation $\hat{U}$, at time $t$. We thus conclude, according to the definition \eqref{P}, that
\eq{
G_{ij}=\frac{\hbar}{4\pi}\left(\delta_{ij}-\frac{1}{2}{\rm Tr}\left[\hat{\mathfrak{r}}\hat{R}_{ii'}\hat{\sigma}_{i'}\hat{\mathfrak{r}}^\dagger\hat{R}_{jj'}\hat{\sigma}_{j'}\right]\right)\,,
}
or in matrix form,
\eq{
\hat{G}=\frac{\hbar}{4\pi}\hat{R}\hat{\mathfrak{g}}\hat{R}^{-1}\,,
\label{G}}
where
\eq{
\mathfrak{g}_{ij}\equiv\delta_{ij}-\frac{1}{2}{\rm Tr}\left[\hat{\mathfrak{r}}\hat{\sigma}_i\hat{\mathfrak{r}}^\dagger\hat{\sigma}_j\right]\,.
\label{gij}}
Note that in order to retain only the relevant symmetric part of $\hat{G}$, the matrix $\hat{\mathfrak{g}}$ entering Eq.~\eqref{G} needs to be symmetrized [i.e., $\hat{\mathfrak{g}}\to(\hat{\mathfrak{g}}+\hat{\mathfrak{g}}^T)/2$], which should be understood as implicit in the above definition \cite{SM}.

In the simplest case of a collinear (ferro-, antiferro-, or ferri-)magnet with the magnetic order oriented along the $z$ axis, the matrix \eqref{gij} simplifies tremendously to
\eq{
\hat{\mathfrak{g}}\to\mathrm{g}_{\rm mix}\{1,1,0\}~~~{\rm (collinear~order)}\,,
\label{gg}}
where $\mathrm{g}_{\rm mix}\equiv1-{\rm Re}\,r^{\uparrow\uparrow}r^{\downarrow\downarrow*}$ is the (real part of the) spin-mixing conductance for a single quantum channel \cite{tserkovPRL02sp}. The $\mathfrak{g}_{zz}$ matrix element is zero as rotations around the $z$ axis commute with the collinear order.

\textit{Multichannel leads.}|It is straightforward to generalize our treatment to an arbitrary number $N$ of transverse quantum channels in the normal-metal lead. In this case, the rotation matrix $\hat{U}$ introduced in Eq.~\eqref{r} should be thought of as $2N\times 2N$ block-diagonal with the usual SU(2) rotations along the diagonal. Repeating our steps, we reproduce Eq.~\eqref{G} for the $3\times3$ dissipative tensor $\hat{G}$, but with the $3\times 3$ matrix $\hat{\mathfrak{g}}$ now given by
\eq{
\mathfrak{g}_{ij}= N\delta_{ij}-\frac{1}{2}\sum_{mn}{\rm Tr}\left[\hat{\mathfrak{r}}_{mn}\hat{\sigma}_i\hat{\mathfrak{r}}_{mn}^\dagger\hat{\sigma}_j\right]\,.
\label{gN}}
Here, $\hat{\mathfrak{r}}_{mn}$ is the $2\times2$ reflection matrix for electrons scattering from channel $n$ into channel $m$, which run from $1\dots N$. As before, a symmetrization with respect to the $i,j$ indices is implicit on the right-hand side of Eq.~\eqref{gN}. This equation, along with Eqs.~\eqref{P}, \eqref{R}, and \eqref{G}, forms a central result of the present work.

For the special case of a collinear order, this again gives Eq.~\eqref{gg}, with the familiar expression for the spin-mixing conductance \cite{tserkovPRL02sp,brataasPRL08}:
\eq{
\mathrm{g}_{\rm mix}= N-{\rm Re}\sum_{mn}r_{mn}^{\uparrow\uparrow}r_{mn}^{\downarrow\downarrow*}~~~{\rm (collinear~order)}\,.
}
In the ferro- or ferrimagnetic cases, this spin-mixing conductance is generically nonzero, so long as electrons experience some exchange upon reflection, which would make $r_{mn}^{\uparrow\uparrow}\neq r_{mn}^{\downarrow\downarrow}$. In the antiferromagnetic case, the spin-mixing conductance is also generally finite, but is dominated by the umklapp scattering channel, in the simplest case of an ideal compensated interface with a translational antiferromagnetic sublattice symmetry \cite{takeiPRB14}.

For a general noncollinear and multichannel case, $\hat{\mathfrak{g}}$ can be diagonalized to yield three non-negative eigenvalues. The corresponding principal axes define a natural magnet-fixed frame of reference for the analysis of the interfacial spin torque and pumping. We can suppose that our laboratory coordinate system is chosen to diagonalize $\hat{\mathfrak{g}}$ (corresponding to the magnetic orientation at $t=0$), with subsequent dynamics yielding a rotated damping tensor \eqref{G}.

\textit{Collinear order.}|Equipped with the (torque-modified) Rayleigh dissipation function \eqref{R}, we can readily construct the boundary conditions for the appropriate magnetic dynamics. To that end, we need to start with the bulk Lagrangian of the magnet. For a collinearly-ordered material, the general (low-temperature) Lagrangian density is given by \cite{auerbachBOOK94}
\eq{
\mathcal{L}=-s\,\mathbf{a}(\mathbf{n})\cdot\partial_t\mathbf{n}+\frac{\chi}{2}\left(\partial_t\mathbf{n}-\gamma\mathbf{n}\times\mathbf{B}\right)^2-\frac{A}{2}(\partial_i\mathbf{n})^2-\mathcal{E}(\mathbf{n})\,,
\label{L}}
where $\mathbf{n}$ is the directional order parameter (s.t., $|\mathbf{n}|\equiv1$), $s$ longitudinal (along $\mathbf{n}$) spin density, $\gamma$ gyromagnetic ratio, $\mathbf{B}$ magnetic field, $A$ order-parameter stiffness, index $i$ runs over spatial (Cartesian) coordinates, $\chi$ is related to the transverse (to $\mathbf{n}$) spin susceptibility, and $\mathcal{E}(\mathbf{n})$ is the local energy density, including anisotropies and Zeeman coupling $-\gamma s\mathbf{n}\cdot\mathbf{B}$ to the longitudinal magnetic moment. $\mathbf{a}(\mathbf{n})$ is a vector potential produced on a unit sphere by a magnetic monopole of unit charge. Antiferromagnets correspond to $s=0$, while low-frequency dynamics in ferro- and ferrimagnets can be obtained by setting $\chi\to0$. 

The Euler-Lagrange  equation of motion is then given by
\eq{
\partial_\nu\frac{\partial\mathcal{L}}{\partial(\partial_\nu\mathbf{n})}-\frac{\partial\mathcal{L}}{\partial\mathbf{n}}+\frac{\partial \mathcal{R}}{\partial(\partial_t\mathbf{n})}=0\,,
\label{EL}}
where $\nu$ runs over all space-time coordinates. $\mathcal{R}\equiv R\,\delta(x)$ should be understood as the spatial density of the Rayleigh dissipation function \eqref{R}, with $R$ here defined \textit{per unit area} of the interface placed at $x=0$ (with the magnet corresponding to $x>0$; see Fig.~\ref{fig1}). For the case of a collinear order,
\eq{
R=\frac{\hbar g_{\rm mix}}{8\pi}\left(\partial_t\mathbf{n}-\boldsymbol{\mu}\times\mathbf{n}/\hbar\right)^2~~~{\rm (collinear~order)}\,,
}
where $g_{\rm mix}$ is the interfacial spin-mixing conductance per unit area. Using Lagrangian \eqref{L}, we find for the equation of motion (taking care to respect the constraint $|\mathbf{n}|\equiv1$):
\eq{
\partial_t(s\mathbf{n}+\mathbf{m})-\gamma\mathbf{m}\times\mathbf{B}-\mathbf{n}\times\left(A\partial_i^2\mathbf{n}-\partial_\mathbf{n}\mathcal{E}\right)=\boldsymbol{\tau}\delta(x)\,,
\label{eom}}
where $\mathbf{m}\equiv\chi\mathbf{n}\times(\partial_t\mathbf{n}-\gamma\mathbf{n}\times\mathbf{B})$ is an auxiliary variable corresponding physically to the transverse spin density (obtained from $\partial_\mathbf{B}\mathcal{L}/\gamma$, which corresponds also to the generators of rotations dictated by the Lagrangian \eqref{L} \cite{andreevSPU80}). The net spin density is thus given by $s\mathbf{n}+\mathbf{m}$. The right-hand side,
\eq{
\boldsymbol{\tau}\equiv\frac{\hbar g_{\rm mix}}{4\pi}\mathbf{n}\times\left(\boldsymbol{\mu}\times\mathbf{n}/\hbar-\partial_t\mathbf{n}\right)\,,
\label{tau}}
is understood as the dissipative torque (spin-current density) produced by the electrons scattering off the interface. Equations \eqref{eom} and \eqref{tau} reproduce and connect the standard ferromagnetic \cite{tserkovRMP05} and antiferromagnetic \cite{baltzCM16} limits (corresponding respectively to setting $\mathbf{m}\to0$ and $s\to0$). Integrating the equation of motion \eqref{eom} near the interface, we finally get
\eq{
-A\mathbf{n}\times\partial_x\mathbf{n}=\boldsymbol{\tau}\,,
}
reflecting the spin continuity at the interface \cite{Note2}. The work done by the torque \eqref{tau}, per unit area and time, is
\eq{
\partial_tw\equiv\boldsymbol{\tau}\cdot\mathbf{n}\times\partial_t\mathbf{n}=\frac{\hbar g_{\rm mix}}{4\pi}\left(\boldsymbol{\mu}\times\mathbf{n}/\hbar-\partial_t\mathbf{n}\right)\cdot\partial_t\mathbf{n}\,.
}
The second term, $\propto-(\partial_t\mathbf{n})^2$, here, is just the ordinary Gilbert damping endowed by the metallic reservoir \cite{tserkovPRL02sp}. The first term reflects the antidamping nature of the spin-transfer torque, for the appropriate orientation of the spin accumulation.

\textit{Spin glass.}|We consider now the opposite extreme of a disordered magnet, in which the orientation of the individual spins are randomly distributed due to frustrated exchange interactions. The full SO(3) group of spin rotations is broken in the ground state, characterized by a matrix or Edwards-Anderson-like order parameter \cite{EA}. Slow (in a hydrodynamic sense) deviations from equilibrium are represented by a vector $\boldsymbol{\theta}=\left(\theta_x,\theta_y,\theta_z\right)$ of rotation angles along the principal axes of $\hat{G}$ defining the laboratory frame. The linearized dynamics is captured by the Lagrangian density \cite{halperinPRB77,andreevSPU80,SM}
\eq{
\mathcal{L}=\frac{\chi}{2}\left(\partial_t \boldsymbol{\theta}+\gamma\mathbf{B}\right)^2+\frac{\chi}{2}\partial_t\boldsymbol{\theta}\cdot\left(\gamma\mathbf{B}\times\boldsymbol{\theta}\right)-\frac{A}{2}\left(\partial_i\boldsymbol{\theta}\right)^2-\mathcal{E}(\boldsymbol{\theta})\,.
\label{Lglass}}
In the absence of anisotropies and net magnetization at equilibrium ($\mathbf{B}=0$), Eq.~\eqref{Lglass} predicts 3 independent polarizations of spin waves with a linear dispersion \cite{foot_m}. 

For a macroscopically isotropic spin configuration, we expect $\hat{G}\propto\hat{1}$ in the presence of an exchange-dominated coupling with the normal-metal electrons. The linearized Rayleigh function (per unit area of the interface) then reads (at $\boldsymbol{\theta}\to0$)
\eq{
R=\frac{\hbar g}{8\pi}\,\left(\partial_t\boldsymbol{\theta}-\frac{\boldsymbol{\mu}}{\hbar}\right)^2~~~{\rm (spin~glass)}\,,
\label{Rglass}}
where $g\equiv g_1=g_2=g_3$ are the eigenvalues of $\hat{\mathfrak{g}}$ in Eq.~\eqref{gN}, normalized by the area. The equation of motion (for a static $\mathbf{B}$) reduces to
\eq{
\partial_t\mathbf{m}-\gamma\mathbf{m}\times\mathbf{B}-A\partial_i^2\boldsymbol{\theta}+\partial_{\boldsymbol{\theta}}\mathcal{E}=\frac{\hbar g}{4\pi}\left(\frac{\boldsymbol{\mu}}{\hbar}-\partial_t\boldsymbol{\theta}\right)\delta\left(x\right)\,,
\label{eomSG}}
where $\mathbf{m}\approx\chi(\partial_t\boldsymbol{\theta}+\gamma\mathbf{B})$ is the spin density ($\equiv\partial_\mathbf{B}\mathcal{L}/\gamma$). As before, this may be interpreted as a continuity equation for spin flow, subject to local precession and interfacial spin transfer. Notice that the pairs $\left(\theta_{\alpha},m_{\alpha}\right)$ are canonically conjugate, a consequence of the fact that the spin-density components define generators of the infinitesimal rotations in the magnetic system. Integrating Eq.~\eqref{eomSG} near the interface leads to the spin-flux continuity at the interface:
\eq{
-A\partial_x\boldsymbol{\theta}=\frac{\hbar g}{4\pi}\left(\frac{\boldsymbol{\mu}}{\hbar}-\partial_t\boldsymbol{\theta}\right)\,.
\label{continuity}
}

\begin{figure}[t]
\includegraphics[width=0.9\columnwidth]{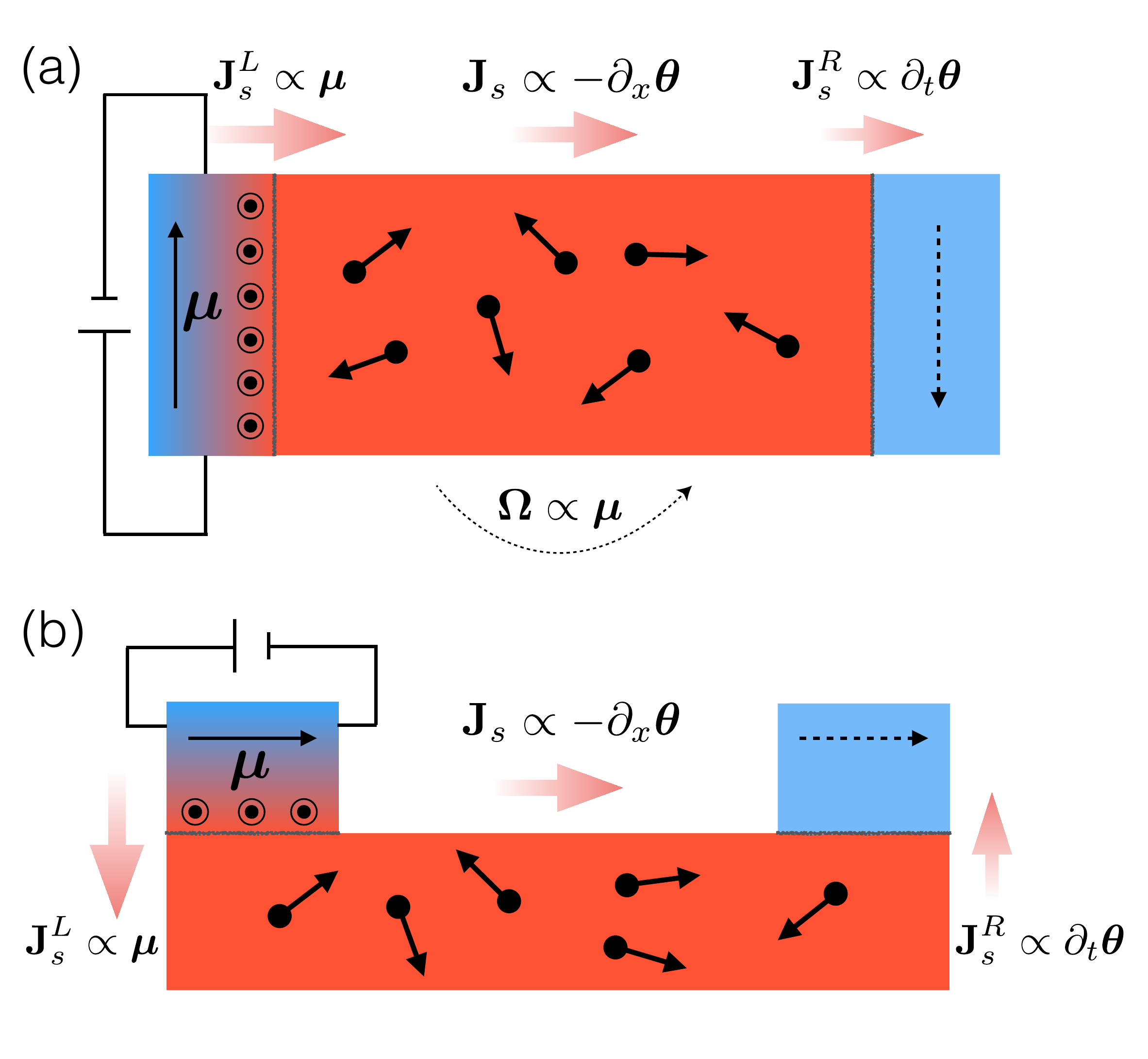}
\caption{Schemes for lateral (a) and vertical (b) spin injection/detection in electrically insulating spin glasses. The signal is sustained by the coherent precession of randomly oriented spins, triggered by the spin accumulation $\boldsymbol{\mu}$ in the left terminal and collected in the right terminal by the reciprocal pumping effect. The steady-state solution for the precession angle about $\boldsymbol{\mu}$ reads $\theta\left(t,x\right)=\Omega t+\theta\left(x\right)$, where $-A\partial_x\theta$ corresponds to the spin-current density in the bulk of the magnet. The (minus) divergence, $A\partial^2_x\theta$, of the spin current in the bulk of the magnet balances the spin damping rate, $\alpha s \Omega$, $\alpha$ being the Gilbert-damping constant and $s$ the high-field saturated spin density. The precession frequency $\Omega$ is proportional to the nonequilibrium spin density along $\boldsymbol{\mu}$ and is determined by the boundary condition in Eq.~\eqref{continuity}. The measured electrical drag signal is negative in (a) and positive in (b), and would follow the numerical estimates of Ref.~\cite{takeiPRL14}.}
\label{fig2}
\end{figure}

This generalized phenomenology enables the study of spin signals transmitted through disordered magnets, which can be probed in a set-up like the one shown in Fig.~\ref{fig2}. The spin accumulation $\boldsymbol{\mu}$ induced by the spin Hall effect in one of the metals triggers the coherent precession of randomly oriented spins in the glass phase, while the signal is collected in a second terminal by means of the reciprocal pumping effect. The steady-state precession frequency $\boldsymbol{\Omega}=\partial_t\boldsymbol{\theta}$ is proportional to the nonequilibrium spin density, $\chi\boldsymbol{\Omega}$, induced in the system. In the geometry of Fig.~\ref{fig2}(a), the frequency is easily obtained \cite{takeiPRL14} by balancing the boundary conditions \eqref{continuity} with the bulk Gilbert damping: $\hbar\boldsymbol{\Omega}=\boldsymbol{\mu}/(2+4\pi\alpha sL/\hbar g)$. In the absence of anisotropies, the signal decays only algebraically with the distance between the terminals $L$, due to the bulk damping $\alpha$, in contrast to the (thermal) spin waves in a collinear magnet \cite{magnon_drag}. Spin glasses provide a (potentially) more versatile platform for long-ranged signal transmission, in comparison to a spin-superfluid state in easy-plane magnets \cite{takeiPRL14}. In particular, they offer flexibility regarding the spin injection and detection geometries, as illustrated in Fig.~\ref{fig2}.

\textit{Discussion.}|The key element of the theory is the modified Rayleigh dissipation function \eqref{R}, which captures the effects of both the spin pumping into the metal reservoir and spin torque by its spin accumulation. The former is directly linked to the dissipation of energy into the normal lead, while the latter to the work on the magnetic dynamics by the spin-polarized electrons. When the spin accumulation $\boldsymbol{\mu}$ exceeds the natural precession frequency $\hbar\boldsymbol{\omega}$, this work can effectively reverse the damping, potentially leading to magnetic instabilities and self-oscillations \cite{slonczewskiPRB89,ralphJMMM08}. (Additional bulk damping of the material would  raise the threshold for such instabilities.) In general, the spin accumulation $\boldsymbol{\mu}$ needs to be calculated self-consistently with the spin-current density $\mathbf{j}_s=-\boldsymbol{\tau}$ flowing into the normal metal.

While our focus has been on electrically insulating magnets, a generalization to conducting magnets is possible by considering transmission as well as reflection of electrons \cite{tserkovPRL02sp}. For the case of sufficiently thick magnets, however, the transmission can generally be expected to lead to a full dephasing of spin transport \cite{tserkovRMP05}, bringing us back to Eq.~\eqref{gN}, which is governed by the reflection coefficients only. Finally, we remark that through Eqs.~\eqref{P} and \eqref{R} we invoked only the dissipative coupling between the magnet and the normal-metal reservoir. Such dissipative spin transfer is known to be the most prominent interfacial process for collinear ferromagnets \cite{tserkovRMP05,brataasPRL08} and antiferromagnets \cite{takeiPRB14}, which is responsible for dynamic instabilities \cite{ralphJMMM08}, thermal-magnon and superfluid spin injection \cite{takeiPRL14,takeiPRB14}, as well as the spin Seebeck physics triggered by heat biases \cite{bauerNATM12,*hoffmanPRB13}. We expect this to naturally extend to the noncollinear case. The nondissipative coupling, which is quantified through the \textit{imaginary} part of the spin-mixing conductance in the collinear case, can be formally captured by redefining the effective Lagrangian (or, equivalently, Hamiltonian or free energy) of the coupled system and renormalizing the reactive coupling coefficients \cite{tserkovRMP05}. While it is in principle possible to account for this both phenomenologically and microscopically in the scattering-matrix formalism \cite{brataasPRB11}, it is beyond our immediate interests. Future works should also address generalizations of our theory to nonrigid exchange dynamics in soft magnets, which may also pump spin and contribute to dissipation, and the role of strong spin-orbit interactions at the interface.

\begin{acknowledgments}
We are grateful to Se Kwon Kim and Pramey Upadhyaya for insightful discussions. This work was supported by the U.S. Department of Energy, Office of Basic Energy Sciences under Award No.~DE-SC0012190.

\end{acknowledgments}


\begin{thebibliography}{29}
\bibitem [{Note4()}]{foot_m}%
  \BibitemOpen
  \bibinfo {note} {In the presence of a field saturated spin density $\mathbf{s}=\chi\gamma\mathbf{B}$,  the linear spin-wave mode polarized along $\mathbf{s}$ survives, $\omega_{\parallel}=\sqrt{A/\chi}\left|\mathbf{k}\right|$, whereas the two remaining modes are hybridized, leading to a gapped, $\omega_{\perp}^+\sim\left|\mathbf{s}\right|/\chi$, and a quadratically dispersing, $\omega_{\perp}^-\sim A\left|\mathbf{k}\right|^2/\left|\mathbf{s}\right|$, branches, similarly to a ferrimagnet. The long-ranged spin transmission proposed in Fig.~\ref{fig2} remains when $\boldsymbol{\mu}$ is aligned with $\mathbf{s}$ (sustaining also a possible collinear anisotropy along the same direction).}\BibitemShut {Stop}%
\end{thebibliography}
\end{document}